# Controlling Surface Plasmons Through Covariant Transformation of the Spin-Dependent Geometric Phase Between Curved Metamaterials

Fan Zhong, [1] Jensen Li, [2] Hui Liu,*[1] and Shining Zhu[1]

Corresponding email: liuhui@nju.edu.cn

[1]National Laboratory of Solid State Microstructures & School of Physics, Collaborative Innovation Center of Advanced Microstructures, Nanjing University, China
[2]School of Physics and Astronomy, University of Birmingham, UK

## Abstract

General relativity uses curved space-time to describe accelerating frames. The movement of particles in different curved space-times can be regarded as equivalent physical processes based on the covariant transformation between different frames. In this work, we use one-dimensional curved metamaterials to mimic accelerating particles in curved space-times. The different curved shapes of structures are used to mimic different accelerating frames. The different geometric phases along the structure are used to mimic different movements in the frame. Using the covariant principle of general relativity, we can obtain equivalent nanostructures based on space-time transformations, such as the Lorentz transformation and conformal transformation. In this way, many covariant structures can be found which produce the same surface plasmon fields when excited by spin photons. A new kind of accelerating beam, the Rindler beam, is obtained based on the Rindler metric in gravity. Very large effective indexes can be obtained in such systems based on geometric phase gradient. This general covariant design method can be extended to many other optical media.

About a hundred years ago, Albert Einstein told us that the space-time continuum should be understood as a curved geometric space. The equivalence between space-times from different observers relies on the fact that the form of the governing equations remains the same. A uniformly accelerating frame is intrinsically the same as an inertial frame with gravity. This recognition allows the inclusion of gravity to generalize the Minkowski space-time in special relativity, allowing a global geometric description of space-time with gravitational masses. Such a geometric theory of gravity successfully predicted gravitational lensing [1], black holes [2,3], and gravitational waves [4]. Surprisingly, a geometric understanding of space-time can be used in optical designs. In 2006, Pendry and Leonhardt proposed a transformation approach to design an invisibility cloak [5,6]. By writing the form-invariant Maxwell's equations in vacuum (the "inertial" frame) with the transformed coordinates in another space [7-9], the equivalence between the two spaces induces generally inhomogeneous indices of refraction filling up the transformed space (i.e. the "gravity" in a "non-inertial" frame), which are then realized by metamaterials [10-18]. Invisibility carpet cloaks [19-21], illusion optics [22], Talbot effects [23] and further extensions to acoustics [24-27], elastic waves [28-30], thermal control [31-33] and even matter waves [34] have been developed. Interestingly, these inhomogeneous indices can then be used to explore different general-relativistic phenomena, like particle motion around a gravitational field, whose principles can now be easily experimentally studied in desktop-scale experiments [35-46]. Besides transformation optics (TO), some other optical structures, such as surfaces of revolution [47,48] and optical lattices [49,50], were also recently reported to mimic general relativity.

When the geometric interpretation is carried from space-time to the state-space of light, the spin-orbit interaction (SOI) can be understood in a geometric-phase setting. An optical spin-Hall effect was thus predicted and directly observed in experiments [51,52]. It allows the usage of metasurfaces [53-61] to implement spin-enabled optical devices using an inhomogeneous profile of metamaterial atoms providing the necessary SOI [62-71]. The incorporation of SOI into the inhomogeneous material profile also provides a handy tool to study the specific motions of particles. Recently, P. Genevet et al. have utilized surface plasmon polaritons (SPPs) generated on a metasurface through SOI as an analogous system to study Cherenkov radiation from a charged particle moving with a constant high speed [72]. A chain of metamaterial atoms lying on a straight line on the metasurface is taken as the analogous particle motion along the same trajectory with constant velocity, which is represented by the spatial rate of change of geometric phase in these metamaterial atoms.

As P. Genevet's paper described [72], in the excitation process of surface plasmons, geometric phase can be employed to mimic different movements in a straight line. In that paper, only geometric phase is taken into consideration, and the effects of curved shapes of structures is not included. On the other hand, as some recent papers proposed, curved shape also plays an important role in designing optical and electric devices [47-49,73]. In this work, we propose a novel and general method to combine curved shape and geometric phase together to design optical structures. Here, drawing on special and general relativity, the different curved shapes of structures are related to different coordinate frames. The different geometric phase profiles inside the structures are used to mimic different movements in the same frame. In this way, we can obtain different covariant transformations of geometric phase between nanostructures with different curved shapes, such as Lorentz transformations and conformal transformations. This enables us to find many equivalent nanostructures which produce the same surface plasmon fields. The well-known Rindler metric in general relativity is employed to generate a new kind of plasmonic accelerating beam, the Rindler beam. Besides surface plasmons, the proposed covariant design method can also be extended to other optical structures.

We would like to use a one-dimensional chain of metamaterial atoms to generate a predefined caustic of surface plasmon polaritons (SPP) on a metal surface, schematically shown as the orange curve in Fig. 1(a). A series of nano slots with orientation profile $\theta(l)$ is placed along a chosen trajectory $l$, the green curved line in the figure, where $\theta$ is the angle between the nano slot and the trajectory. When the chain of atoms is illuminated by circularly polarized light, the generated SPP signal carries a geometric phase given by $\Phi = \pm(2\theta - \arg(k_x + ik_y))$ [59]. In our experiments, we use left-handed circularly polarized incidence, and the term $\arg(k_x + ik_y)$ can be neglected: $\Phi = -2\theta$. When the orientation of the atoms changes along the trajectory, a SPP-ray is emitted at an angle $\varphi$ governed by $k_{\text{SPP}} \cos\varphi = d\Phi/dl$. This angle is defined with respect to the direction along the trajectory, as shown in Fig. 1(a). The SPP rays (red) emitted from different positions along the trajectory vary in direction and form an envelope, which is called the caustic. Here, we develop a metric route to obtain the required profile of the geometric phase $\Phi(l)$. For a caustic described by a parametric description $(x, y) = (X(\tau), Y(\tau))$ where $\tau$ is called the caustic parameter, we first promote the caustic parameter to a global function defined for the spatial region below the caustic by associating $\tau$ to a general location $(x, y)$, so that the caustic is constituted from the tangent points of SPP rays parametrized by the same $\tau$ passing through $(x, y)$. Equivalently,

$$\frac{dY(\tau)}{d\tau}(x - X(\tau)) = \frac{dX(\tau)}{d\tau}(y - Y(\tau)). \tag{1}$$

The global $\tau(x, y)$, obtained from the above equation, is a convenient geometrical object representing the structure of the caustic. The contour lines of $\tau(x,y)$ are straight lines, namely the SPP rays constituting the caustic. Figure 1(b) show two such SPP rays, labeled by $\tau$ and $\tau + \Delta\tau$, together with the trajectory of metamaterial atoms. The perpendicular distance can now be written in terms of the global variable $\tau$ as $\Delta\tau/|\nabla\tau|$. The length along the trajectory is $\Delta l$. The wavefront of the SPP-ray is shown by the dashed lines, whose effective elapsed optical path should be the geometric phase $\Delta\Phi/k_{SPP}$. Therefore, we obtain the metric relationship

$$d\tau^2 = |\nabla\tau|^2 (dl^2 - (d\Phi/k_{SPP})^2). \tag{2}$$

For a given caustic and a chosen trajectory of atoms, we can obtain the required geometric phase profile $\Phi(l)$ by integrating the metric relationship along the trajectory as

$$\Phi(l) = k_{SPP} \int \sqrt{1 - (1/|\nabla\tau|)^2 (d\tau/dl)^2} \, dl. \tag{3}$$

Such a geometric phase profile is then converted by the orientation profile of the metamaterial atoms. The formulation of $\tau(x,y)$ gives us huge mathematical convenience. However, we note that the parametrization of the caustic is actually not unique. Different choices of $\tau$ are like different gauges, while the obtained $\Phi(l)$ is gauge-independent. Thus, the construction method we offer provides simplicity, even for complex curved lines.

Moreover, the $\tau$ object can exist even if a caustic does not form. A linear progressing $\tau(x,y)$, along a fixed direction, can be used to represent simple planar radiation. In such a case, if we choose any straight-line trajectories (e.g. the trajectory $l$ in Fig. 1(c)), it corresponds to a chain of metamaterial atoms with linear geometric phase/orientation profile. This chain radiates SPPs at a fixed angle $\varphi$ (again with $\cos\varphi = (1/k_{SPP}) d\Phi/dl$). Such radiation can be interpreted as an analogue of Cherenkov radiation from a charged particle moving at a constant velocity larger than the speed of light in a dielectric medium; see, for example, P. Genevet's work [72]. In Fig. 1(c), we also draw another straight-line trajectory $l'$, which crosses the SPP-rays at a different angle $\varphi'$. The two trajectories are equivalent in generating SPP-rays, with the orientation of metamaterial atoms changing at two different but constant rates. The equivalence can be interpreted as a conservation of $\Delta\tau$ between the two trajectories: $\Delta\tau^2 = dl^2 - (d\Phi/k_{\text{SPP}})^2 = dl'^2 - (d\Phi'/k_{\text{SPP}})^2$. The transformation between the two equivalent trajectories can be written as

$$|\nabla\tau|\begin{pmatrix} dl \\ d\Phi/k_{\text{SPP}} \end{pmatrix} = \begin{pmatrix} \cosh\zeta & \sinh\zeta \\ \sinh\zeta & \cosh\zeta \end{pmatrix} |\nabla\tau|' \begin{pmatrix} dl' \\ d\Phi'/k_{\text{SPP}} \end{pmatrix}, \quad (4)$$

where $\zeta$ is called the rigidity between the two trajectories. Here, if the two meta-chains are regarded as two different reference frames, Eq. (5) can be seen as the Lorentz transformation between them [75], under the analogue's mapping $t_M = |\nabla\tau|l, x_M = |\nabla\tau|\Phi/k_{\text{SPP}}$,

$$\begin{pmatrix} dt_M \\ dx_M \end{pmatrix} = \begin{pmatrix} \cosh\zeta & \sinh\zeta \\ \sinh\zeta & \cosh\zeta \end{pmatrix} \begin{pmatrix} dt'_M \\ dx'_M \end{pmatrix}, \quad (5)$$

which guarantees the conservation of metric $d\tau^2 = dt_M^2 - dx_M^2 = dt'^2_M - dx'^2_M$.

We now extend our scheme to more general situations. For more general motion, although the velocity changes either in amplitude or in direction, we can still use the same strategy to simulate the motion on a straight or curved line as previously described. Under this circumstance, the appearance of the caustic, which is the envelope of nonparallel $\tau$ lines [74], is characteristic of "Bremsstrahlung radiation", and the emergence of this radiation requires acceleration. For experimental implementation, the change of $\nabla\Phi$ needs to be slow so that, when chains of meta-atoms share the same family of SPP rays ($\tau$ lines), the transformation of trajectories is established as shown in Fig. 2(a).

Here, we show an example of transformation between a motion of constant velocity and an accelerating motion by mapping $l$ to "time" and $\Phi/k_{\text{spp}}$ to "length" in Fig. 2(a). One of the trajectories, marked in white with a dashed horizontal line, has its acceleration set to be proportional to $l_1^{-3/2}$. The other trajectory, in green with a curved dashed line, has constant velocity (see Fig. 5s in the Supplemental Material [75]). The corresponding SEM (scanning electron microscope) pictures of the metallic structures are shown in Figs. 2(b, e) for $l_1$ and $l_2$. Simulations are shown in Figs. 2(c, f), and our experiments are shown in Figs. 2(d, g). All four results have the same caustic, indicating shared $\tau$ lines, labeled by green or white dashed lines. This provides a good match between theory and experiment. Also, the fact that another caustic appears in the test sample due to the SPP signal, without carrying geometric phase, provides a clue to the trajectory's curvature. Thus, we extend our experimental platform to curved chains for exploring the more general situation.

To explore the transformation in the more general situation, we need to establish a curved space-time picture based on inhomogeneous distributed $\tau$ lines through analogy to general relativity. Thus, we map $t_C = l$ and $x_C = \Phi/k_{\text{spp}}$ and assume that the two-dimensional space-time represented by this trajectory has the mathematic form

$$d\tau^2 = g(t_C, x_C)^2 (dt_C^2 - dx_C^2). \tag{6}$$

The metric here covers the more general situation and indicates the curvature effect. Since we have established the space-time on trajectories, we might wonder whether the motions defined on different trajectories are in fact the same motion, as shown in Fig. 3. If on two trajectories the motions are the same, then in different space-times, the metric should obey $g_{\mu\nu} = (\partial x^i/\partial x^\mu)(\partial x^k/\partial x^\nu) g_{ik}$. This maintains the covariance of the metric, if $\varphi$ remains the same (see the Supplemental Material [75]). Thus, the transformation can be divided into two steps. The first step is applying the conformal transformation, that $|\nabla \tau| \to |\nabla \tau|'$ with $\varphi = \varphi'$,

$\cosh \zeta = 1$ and $\sinh \zeta = 0$, corresponding to expanding or shrinking the local length of the trajectory. The second step is using the Lorentz transformation, corresponding to rotating the local trajectory element. Therefore, the more general transformation is

$$|\nabla\tau|(l)\begin{pmatrix}dt_C\\dx_C\end{pmatrix} = |\nabla\tau|(l')\begin{pmatrix}\cosh\zeta & \sinh\zeta\\\sinh\zeta & \cosh\zeta\end{pmatrix}\begin{pmatrix}dt'_C\\dx'_C\end{pmatrix}, \quad (7)$$

which contains conformal covariance. Here there is an approximation that, due to slowly varying $\varphi$, we can locally apply Lorentz transformations. Now, the general transformation is established between two curved spaces in our scheme, which can be exploited to explore the transformation of space-times and the corresponding phenomena in general relativity.

When studying the close neighborhood of a black hole, Rindler coordinates are very useful in describing the geometry close to the event horizon with uniform acceleration [76]. A specific experiment result is now presented, analogous to the Rindler transformation in our established scheme. In Fig. 4(a), we first define the motion marked by white dashed horizontal line to be accelerating to the right with $d\Phi/k_{\text{spp}}/dl = -\tanh(\beta e^{\alpha l}/\alpha)$. We set $\tau = \varphi(l) + \tan^{-1} dy/dx$ (the value of the angle between the $\tau$ line and $x$ axis) on every $\tau$ line, where $x$ and $y$ describe the position of $l(x,y)$ for numerical calculation. This choice of value leads to $|\nabla\tau|(l_1) = \beta e^{\alpha l_1}$ on the white dashed line, where $\alpha$ and $\beta$ are small constants. Also, as shown in Fig. 4(a), on the cyan line $|\nabla\tau|(l_2) = \beta_2$, with $\beta_2$ a constant, and on the green trajectory the motion has constant speed (see Fig. 7s in the Supplemental Material [75]). Here, a Rindler-analogous transformation occurs between the two trajectories (the white line and the cyan line) with a positive constant $\beta_1$, $\beta_1^2 e^{2\alpha l_1}(dl_1^2 - d\Phi_1^2/k_{\text{spp}}^2) = dl_2^2 - d\Phi_2^2/k_{\text{spp}}^2$, which implies $|\nabla\tau|(l_1)/|\nabla\tau|(l_2) = \beta_1 e^{\alpha l_1}$. We experimentally implemented three situations in our metamaterials platform, with corresponding SEM pictures for $l_1$, $l_2$ and $l_3$ in

Figs. 4(b, e, h). The simulations in Figs. 4(c, f, i) show the same caustic, marked by a green dashed line, that appears in experiments in Figs. 4(d, g, j) highlighted by the white dashed line. Showing the same caustic, which represents "Bremsstrahlung radiation" in the experiment, reveals a good match between theory, simulation and experiment as realization of Rindler-analogous transformations. The caustic obtained based on Rindler metrics is a completely new kind of accelerating beam. It is only one special case of designing new kinds of accelerating beams from the metrics in gravity. In addition, many other metrics in general relativity are also worth trying in future research.

As in the examples mentioned above, when a transformation provides an equivalent curved trajectory of motion within the same $\tau$ plane, there is a gauge freedom on the $\tau$ line. If we choose another value $\tilde{\tau}(l) = f(\tau(l))$, the gauge-independent quantity between two coordinates is $g(t_{C1})/g(t_{C2}) = \tilde{g}(t_{C1})/\tilde{g}(t_{C2})$. As in the equivalence principle in general relativity, we cannot distinguish a specific coordinate system due to the gauge freedom, but the comparison is meaningful, and this ratio actually modifies the relation between these two different space-times. As in the examples implementing "Bremsstrahlung radiation", if we change the choice of value $\tau$ to $\tilde{\tau}$, their ratio still has $g(l_1)/g(l_2) = \tilde{g}(l_1)/\tilde{g}(l_2)$ (see Fig. 9s in the Supplemental Material [75]). This gauge-independence shows invariance in our scheme.

In summary, we propose a very flexible method to design transformation optical structures through combining curved shape and geometric phase. In this method, gauge freedom leads us to find many equivalent structures through covariant transformations, structures which share the same optical functions. Some interesting metrics in general relativity are employed for the first time to generate a new kind of plasmonic accelerating beam, the Rindler beam.

Up to now, different methods and structures in metamaterials have been used to mimic curved space [35-46,77,78]. Each kind of metamaterials has both advantages and limitations. Usually, it is not easy to make the effective index of metamaterials satisfy the requirement of curved space. Sometimes, the effective parameters need to be very large. This work provides an easy way to obtain a large effective index based on the geometric phase of spin photons. If we define the speed of a photon along the meta-chain as $v = (d\Phi/dl)/k_{\text{spp}}$, we can define the effective index of a structure as $n_{eff} = c/v = c \cdot k_{\text{spp}}/(d\Phi/dl)$. Here, as the geometric phase $\Phi$ is only determined by the nano slot rotation angle $\Phi = \pm 2\theta$ (here "$\pm$" represents geometric phases of different optical spins), we can change the refractive index $n_{eff}$ by controlling $\theta$. As the sign and amplitude of $\theta$ can be conveniently tuned, the effective index obtained in this way can be any value $n \in (-\infty, \infty)$ without being limited by resonance dispersions or the intrinsic properties of constitutive materials (see supplementary figure 5s (e-f) and figure 7s (g-i) [75]). In experiments, we only need to tune the rotation angle of nano slots. It is very easy to implement based on present techniques without need of new materials or new complex structure design. Another advantage of our work is that photonic spin can provide more freedom to control the effective index. As the sign of geometric phase is determined by incident spin, we can easily change the sign of the effective index by just reversing the photon spin. Furthermore, our method is not limited to surface plasmons, as it can also be extended to dielectric waveguides if we use dielectric particles to obtain geometric phase instead of metallic nano slots.

This work was financially supported by National Key R&D Program of China (2017YFA0303702 and 2017YFA0205700), the National Natural Science Foundation of China (11690033, 11621091, and 61425018).

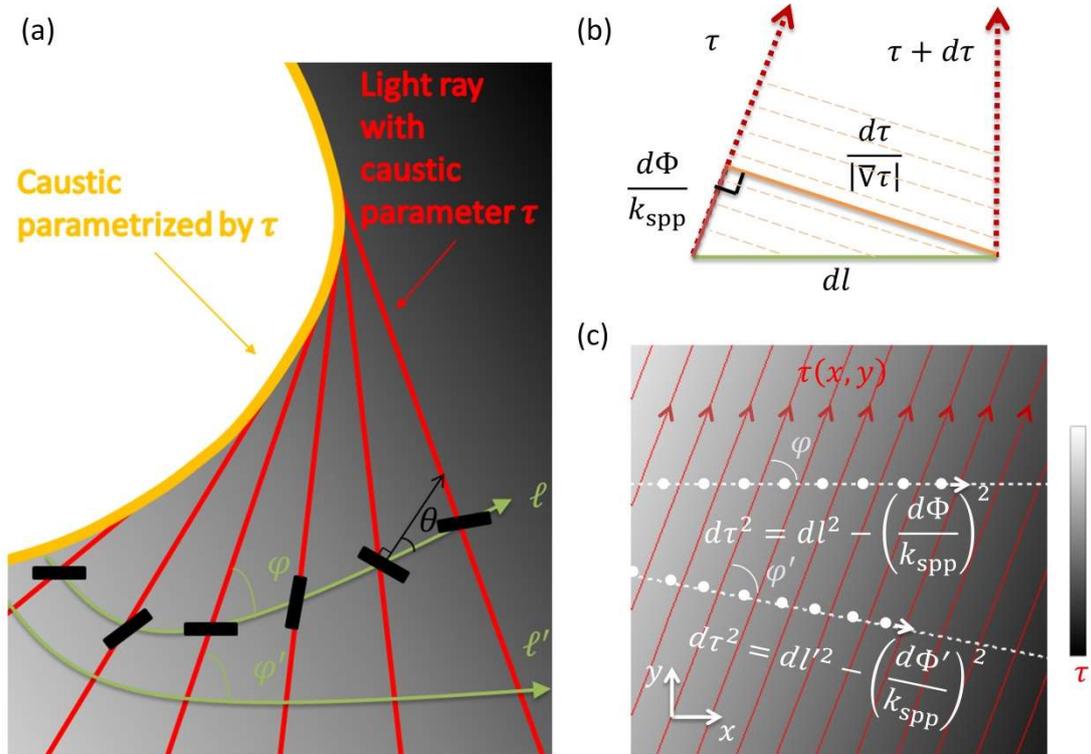

FIG. 1. (a) the transformation sharing same caustic. (b) the geometric relationship between $\tau$, $\Phi$, and $l$. (c) the transformation that occurs on the caustic through Cherenkov radiation, which can be mapped to special relativity.

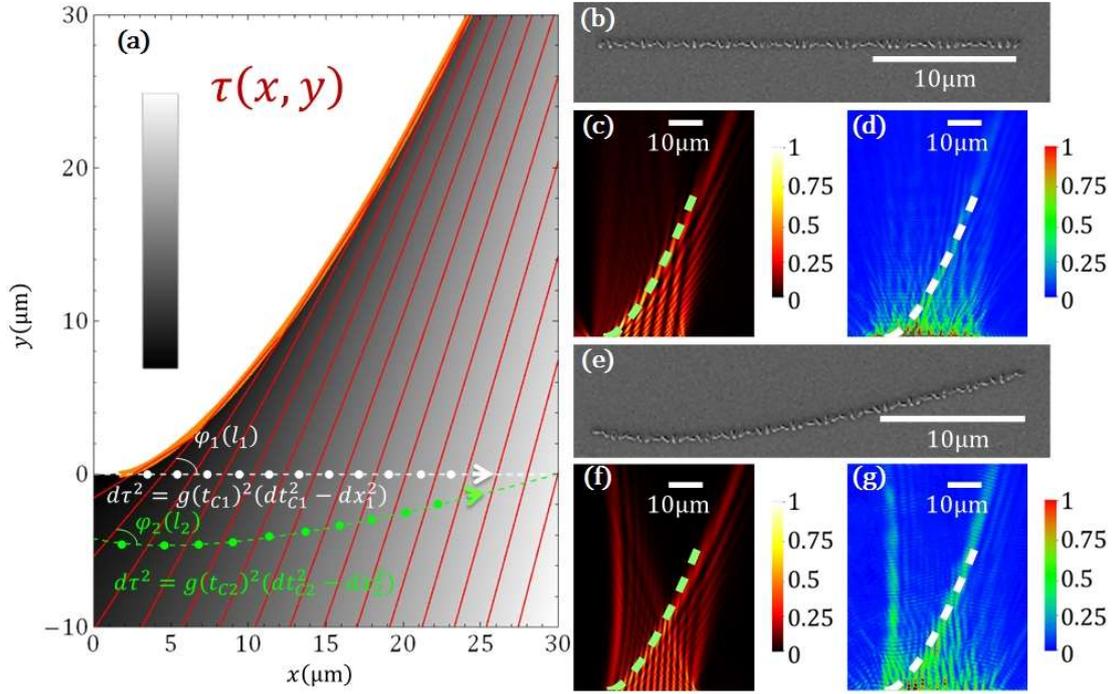

FIG. 2. Transformations. (a) the theoretic calculation. SPP rays marked in red lead to a caustic marked in orange. A transformation occurs between the white dashed line and the green dashed. (b-d) figures for the case of the white dashed line. (e-g) for the case of the green dashed line. (b, e) SEM pictures of our samples. (c, f) full wave simulations by COMSOL with the green dashed line labeling the corresponding caustic. (d, g) experimental results with the white dashed line labeling the corresponding caustic. Legends on the right part of simulations and experiments mark the normalized amplitude with different colors.

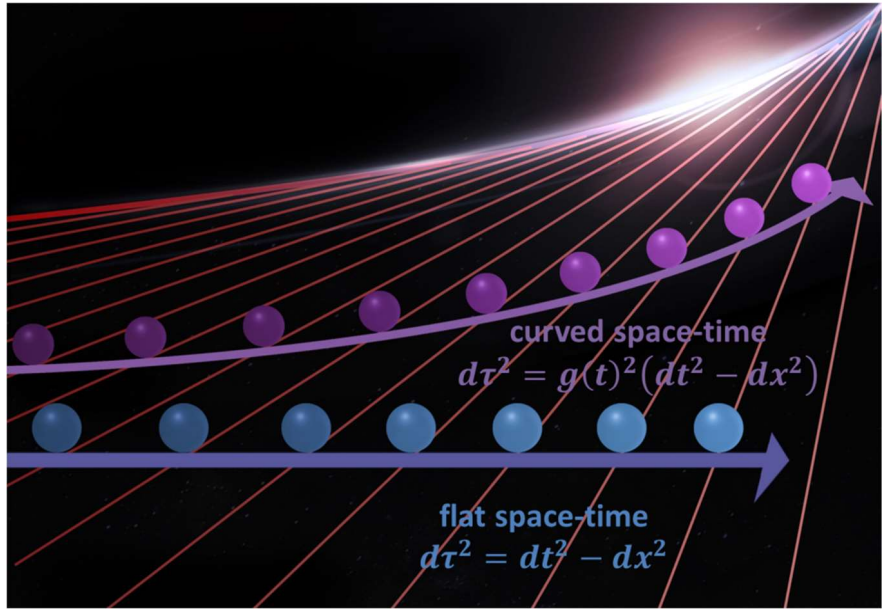

FIG. 3. Geometric picture describing the same event in different coordinates through mimicking Bremsstrahlung radiation of moving particles (blue ball and purple ball) in flat space-time and curved space-time respectively.

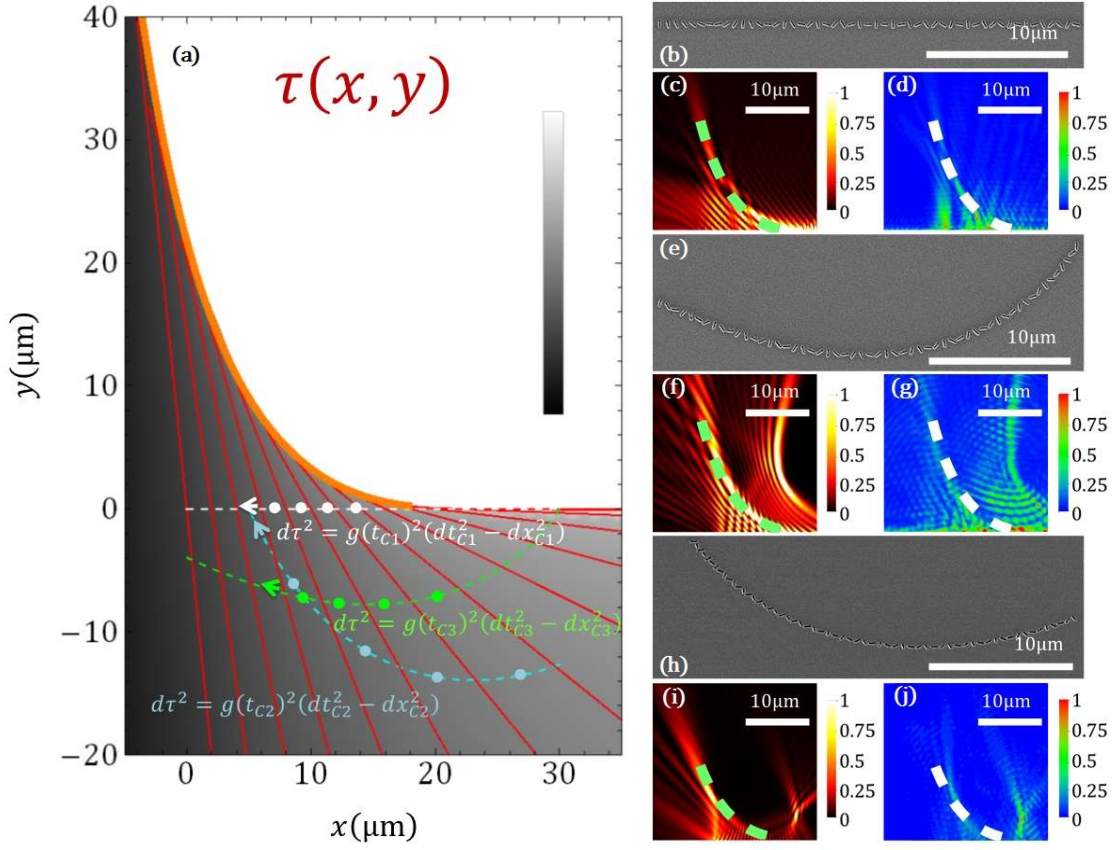

FIG. 4. Transformations analogous to the Rindler transformation in general relativity. (a) shows SPP rays in red and caustic in orange. The dashed white line, green line and cyan line of nano slots represent different corresponding coordinates. (b, e, h) SEM pictures of samples of (b) white dashed line, (e) green dashed line and (h) cyan dashed line with inclination angle of $52°$. (c, d, f, g, i, j) are the simulations and experiments of these three cases. (c, f, i) are full wave simulations by COMSOL with the green dashed line labeling the corresponding caustic. (d, g, j) are experimental results with the white dashed line labeling the corresponding caustic. Legends on the right part of simulations and experiments describe the normalized amplitude with different colors.

# References

[1] A. Einstein, Science **84**, 506 (1936).
[2] N. I. Shakura and R. A. Sunyaev, Astron. Astrophys. **24**, 337 (1973).
[3] M. Bañados, C. Teitelboim, and J. Zanelli, Phys. Rev. Lett. **69**, 1849 (1992).
[4] B. P. Abbott *et al.*, Phys. Rev. Lett. **116**, 061102 (2016).
[5] U. Leonhardt, Science **312**, 1777 (2006).
[6] J. B. Pendry, D. Schurig, and D. R. Smith, Science **312**, 1780 (2006).
[7] T. T. Robert, S. A. Cummer, and J. Frauendiener, J. Opt. **13**, 024008 (2011).
[8] Y. Luo, D. Y. Lei, S. A. Maier, and J. B. Pendry, Phys. Rev. Lett. **108**, 023901 (2012).
[9] J. B. Pendry, Y. Luo, and R. Zhao, Science **348**, 521 (2015).
[10] D. Schurig, J. J. Mock, B. J. Justice, S. A. Cummer, J. B. Pendry, A. F. Starr, and D. R. Smith, Science **314**, 977 (2006).
[11] W. Cai, U. K. Chettiar, A. V. Kildishev, and V. M. Shalaev, Nat. Photonics **1**, 224 (2007).
[12] H. Chen, B. I. Wu, B. Zhang, and J. A. Kong, Phys. Rev. Lett. **99**, 063903 (2007).
[13] G. W. Milton, N.-A. P. Nicorovici, R. C. McPhedran, K. Cherednichenko, and Z. Jacob, New J. Phys. **10**, 115021 (2008).
[14] I. I. Smolyaninov, V. N. Smolyaninova, A. V. Kildishev, and V. M. Shalaev, Phys. Rev. Lett. **102**, 213901 (2009).
[15] T. Ergin and M. Wegener, Science **328**, 337 (2010).
[16] R. C. Mitchell-Thomas, T. M. McManus, O. Quevedo-Teruel, S. A. R. Horsley, and Y. Hao, Phys. Rev. Lett. **111**, 213901 (2013).
[17] O. Quevedo-Teruel, W. Tang, R. C. Mitchell-Thomas, A. Dyke, H. Dyke, L. Zhang, S. Haq, and Y. Hao, Sci. Rep. **3**, 1903 (2013).
[18] J. B. Pendry, P. A. Huidobro, Y. Luo, and E. Galiffi, Science **358**, 915 (2017).
[19] J. Li and J. B. Pendry, Phys. Rev. Lett. **101**, 203901 (2008).
[20] J. Valentine, J. Li, T. Zentgraf, G. Bartal, and X. Zhang, Nat. Mater. **8**, 568 (2009).
[21] M. Gharghi, C. Gladden, T. Zentgraf, Y. Liu, X. Yin, J. Valentine, and X. Zhang, Nano Lett. **11**, 2825 (2011).
[22] Y. Lai, J. Ng, H. Y. Chen, D. Z. Han, J. J. Xiao, Z.-Q. Zhang, and C. T. Chan, Phys. Rev. Lett. **102**, 253902 (2009).
[23] X. Wang, H. Chen, H. Liu, L. Xu, C. Sheng, and S. Zhu, Phys. Rev. Lett. **119**, 033902 (2017).
[24] H. Chen and C. T. Chan, Appl. Phys. Lett. **91**, 357 (2007).
[25] S. A. Cummer and D. Schurig, New J. Phys. **9**, 45 (2007).
[26] S. Zhang, C. Xia, and N. Fang, Phys. Rev. Lett. **106**, 024301 (2011).
[27] P. Zhang, T. Li, J. Zhu, X. Zhu, S. Yang, Y. Wang, X. Yin, and X. Zhang, Nat. Commun. **5**, 4316 (2014).
[28] G. W. Milton, M. Briane, and J. R. Willis, New J. Phys. **8**, 248 (2006).
[29] M. Brun, S. Guenneau, and A. B. Movchan, Appl. Phys. Lett. **94**, 061903 (2009).
[30] N. Stenger, M. Wilhelm, and M. Wegener, Phys. Rev. Lett. **108**, 014301 (2012).
[31] Y. Ma, L. Lan, W. Jiang, F. Sun, and S. He, NPG Asia Mater. **5**, e73 (2013).
[32] T. Han, X. Bai, D. Gao, J. T. L. Thong, B. Li, and C.-W. Qiu, Phys. Rev. Lett. **112**,

054302 (2014).

[33] H. Xu, X. Shi, F. Gao, H. Sun, and B. Zhang, Phys. Rev. Lett. **112**, 054301 (2014).
[34] S. Zhang, D. A. Genov, C. Sun, and X. Zhang, Phys. Rev. Lett. **100**, 123002 (2008).
[35] D. A. Genov, S. Zhang, and X. Zhang, Nat. Phys. **5**, 687 (2009).
[36] E. E. Narimanov and A. V. Kildishev, Appl. Phys. Lett. **95**, 041106 (2009).
[37] H. Chen, R.-X. Miao, and M. Li, Opt. Express **18**, 15183 (2010).
[38] Q. Cheng, T. J. Cui, W. X. Jiang, and B. G. Cai, New J. Phys. **12**, 063006 (2010).
[39] Smolyaninov, II and E. E. Narimanov, Phys. Rev. Lett. **105**, 067402 (2010).
[40] D. A. Genov, Nat. Photonics **5**, 76 (2011).
[41] M. W. McCall, A. Favaro, P. Kinsler, and A. Boardman, J. Opt. **13**, 024003 (2011).
[42] I. I. Smolyaninov, E. Hwang, and E. Narimanov, Phys. Rev. B **85**, 235122 (2012).
[43] C. Sheng, H. Liu, Y. Wang, S. N. Zhu, and D. A. Genov, Nat. Photonics **7**, 902 (2013).
[44] Smolyaninov, II, B. Yost, E. Bates, and V. N. Smolyaninova, Optics express **21**, 14918 (2013).
[45] C. Sheng, R. Bekenstein, H. Liu, S. Zhu, and M. Segev, Nat. Commun. **7**, 10747 (2016).
[46] Y.-L. Zhang, J. B. Pendry, and D. Y. Lei, Phys. Rev. B **96**, 035430 (2017).
[47] V. H. Schultheiss, S. Batz, and U. Peschel, Nat. Photonics **10**, 106 (2015).
[48] R. Bekenstein, Y. Kabessa, Y. Sharabi, O. Tal, N. Engheta, G. Eisenstein, A. J. Agranat, and M. Segev, Nat. Photonics **11**, 664 (2017).
[49] R. Beravat, G. K. Wong, M. H. Frosz, X. M. Xi, and P. S. Russell, Sci. Adv. **2**, e1601421 (2016).
[50] E. Lustig, M.-I. Cohen, R. Bekenstein, G. Harari, M. A. Bandres, and M. Segev, Phys. Rev. A **96**, 041804 (2017).
[51] M. Onoda, S. Murakami, and N. Nagaosa, Phys. Rev. Lett. **93**, 083901 (2004).
[52] O. Hosten and P. Kwiat, Science **319**, 787 (2008).
[53] X. Chen *et al.*, Nat. Commun. **3**, 1198 (2012).
[54] L. Huang, X. Chen, H. Muhlenbernd, G. Li, B. Bai, Q. Tan, G. Jin, T. Zentgraf, and S. Zhang, Nano Lett. **12**, 5750 (2012).
[55] M. Kang, T. Feng, H. T. Wang, and J. Li, Opt. Express **20**, 15882 (2012).
[56] L. Huang, X. Chen, B. Bai, Q. Tan, G. Jin, T. Zentgraf, and S. Zhang, Light Sci. Appl. **2**, e70 (2013).
[57] G. Li, M. Kang, S. Chen, S. Zhang, E. Y. Pun, K. W. Cheah, and J. Li, Nano Lett. **13**, 4148 (2013).
[58] J. Lin, J. P. Mueller, Q. Wang, G. Yuan, N. Antoniou, X. C. Yuan, and F. Capasso, Science **340**, 331 (2013).
[59] S. Xiao, F. Zhong, H. Liu, S. Zhu, and J. Li, Nat. Commun. **6**, 8360 (2015).
[60] K. Y. Bliokh, F. J. Rodríguez-Fortuño, F. Nori, and A. V. Zayats, Nat. Photonics **9**, 796 (2015).
[61] S.-Y. Lee, K. Kim, S.-J. Kim, H. Park, K.-Y. Kim, and B. Lee, Optica **2**, 6 (2015).
[62] G. Biener, A. Niv, V. Kleiner, and E. Hasman, Opt. Lett. **27**, 1875 (2002).
[63] Z. E. Bomzon, G. Biener, V. Kleiner, and E. Hasman, Opt. Lett. **27**, 1141 (2002).
[64] E. Hasman, Z. E. Bomzon, A. Niv, G. Biener, and V. Kleiner, Opt. Commun. **209**, 45 (2002).

[65] E. Hasman, G. Biener, A. Niv, and V. Kleiner, Prog. Opt. **47**, 215 (2005).
[66] K. Y. Bliokh, Y. Gorodetski, V. Kleiner, and E. Hasman, Phys. Rev. Lett. **101**, 030404 (2008).
[67] Y. Gorodetski, N. Shitrit, I. Bretner, V. Kleiner, and E. Hasman, Nano Lett. **9**, 3016 (2009).
[68] N. Dahan, Y. Gorodetski, K. Frischwasser, V. Kleiner, and E. Hasman, Phys. Rev. Lett. **105**, 136402 (2010).
[69] N. Shitrit, I. Bretner, Y. Gorodetski, V. Kleiner, and E. Hasman, Nano Lett. **11**, 2038 (2011).
[70] N. Shitrit, I. Yulevich, E. Maguid, D. Ozeri, D. Veksler, V. Kleiner, and E. Hasman, Science **340**, 724 (2013).
[71] R. C. Devlin, A. Ambrosio, N. A. Rubin, J. P. B. Mueller, and F. Capasso, Science **358**, 896 (2017).
[72] P. Genevet, D. Wintz, A. Ambrosio, A. She, R. Blanchard, and F. Capasso, Nat. Nanotechnol. **10**, 804 (2015).
[73] M. Cariglia, R. Giambò, and A. Perali, Phys. Rev. B **95**, 245426 (2017).
[74] L. Froehly, F. Courvoisier, A. Mathis, M. Jacquot, L. Furfaro, R. Giust, P. A. Lacourt, and J. M. Dudley, Opt. Express **19**, 16455 (2011).
[75] See Supplemental Material for details on the theory of an analogue of relativity, sample fabrication, optical measurements, numerical calculations, and guage theory.
[76] L. C. B. Crispino, A. Higuchi, and G. E. A. Matsas, Rev. Mod. Phys. **80**, 787 (2008).
[77] A. Greenleaf, Y. Kurylev, M. Lassas, and G. Uhlmann, Phys. Rev. Lett. **99** (2007).
[78] T. G. Mackay and A. Lakhtakia, Phys. Lett. A **374**, 2305 (2010).

# Supplementary

## Controlling Surface Plasmons Through Covariant Transformation of the Spin-Dependent Geometric Phase Between Curved Metamaterials

Fan Zhong [1], Jensen Li [2,3], Hui Liu*[1], and Shining Zhu[1]

[1]*National Laboratory of Solid State Microstructures & School of Physics, Collaborative Innovation Center of Advanced Microstructures, Nanjing University, Nanjing, 210093 China*
[2]*School of Physics and Astronomy, University of Birmingham, UK*
[3]*Department of Physics, The Hong Kong University of Science and Technology, Clear Water Bay, Hong Kong, China*

## I. Sample fabrications and measurement setup.

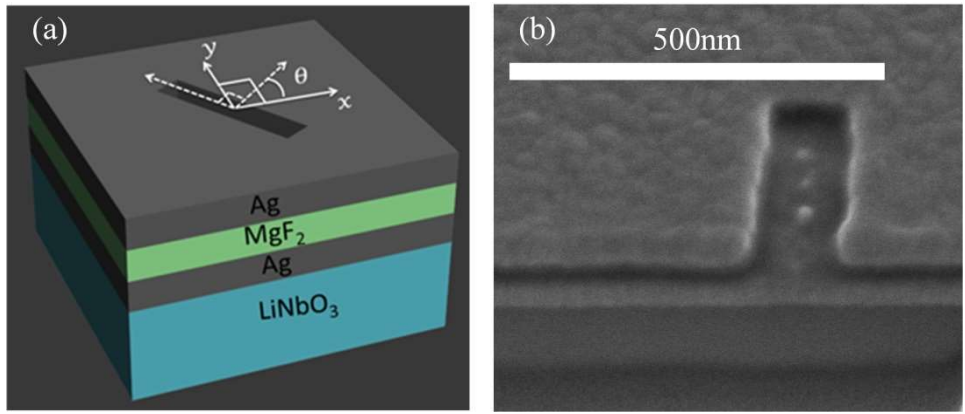

FIG. 1s. Unit cell pictures: (a) is sketch, (b) is SEM picture of cross-section with inclination angle of 52°.

The structure of metamaterial atom and sample picture are given in figure 1s (a) and (b). We fabricated a one-dimensional meta-chain composed of 500nm × 100nm metallic nano-slots on the top layer of a sandwich-like structure through a focused ion beam (FEI Dual Beam HELIOS NANOLAB 600i, 30 keV, 7 pA). This structure, consisting of two silver films separated by a magnesium fluoride spacer, is based on an LN substrate (see Fig. 1s). The top silver film is about 50nm, the magnesium fluoride spacer is about 40nm and the silver film on the LN substrate is about 45nm, which they all are deposited with sputtering-deposition. A semiconductor 1064 nm laser (Coherent Mephisto S500NE) shaped by a slit was used to accurately excite metallic nano-slots with different orientation angles $\theta$, to generate target SPP field on the air-metal surface. An oil-immersed micro-objective associated with the LN substrate was used to collect the SPP signal from the sample's bottom, which was imaged using a high-resolution sCMOS camera. Because of the need for high efficiency when transferring the propagating wave into a surface wave and blocking the transmitted portion, the parameters of metal/dielectric/metal structure were carefully tuned in experiment.

## II. Cherenkov radiation described using concepts in special relativity

In our case, we set $\eta = |\nabla\tau|l$ and $\xi = |\nabla\tau|\Phi/k_{\text{spp}}$. Under the condition $d\xi/d\eta = \cos\varphi$, we will establish a transformation similar to the Lorentz transformation. We start from the line element

$$d\tau^2 = d\eta^2 - d\xi^2 = d\eta'^2 - d\xi'^2 \tag{1s}$$

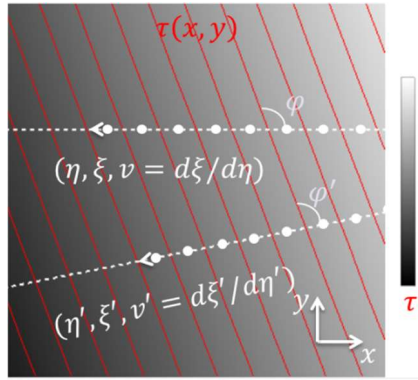

FIG. 2s. Transformation between two Minkowski spaces represented by two white dashed lines.

As in Fig. 2s, for a transformation between two straight lines with different slopes, we use the same $\tau$ line to label the same events. The same motion has different velocity seen by different observers, and this motion specifies an inertial space-time with coordinates $\xi_0$ and $\eta_0$ still satisfying $d\tau^2 = d\eta_0^2 - d\xi_0^2$. Similar to the Lorentz transformation, we introduce an additional parameter $\zeta$ for which

$$\begin{cases} \xi_0 = -\sinh\zeta\,\eta + \cosh\zeta\,\xi \\ \eta_0 = \cosh\zeta\,\eta - \sinh\zeta\,\xi \end{cases} \tag{2s}$$

When we have $d\xi_0 = 0$, we derive $d\xi/d\eta = \tanh\zeta = \cos\varphi$, which is the velocity of the induced Minkowski space-time. In our Cherenkov radiation example:

$$\begin{cases} \xi_0 = -\sinh\zeta\,\eta + \cosh\zeta\,\xi \\ \eta_0 = \cosh\zeta\,\eta - \sinh\zeta\,\xi \end{cases}, \begin{cases} \xi_0 = -\sinh\zeta'\,\eta' + \cosh\zeta'\,\xi' \\ \eta_0 = \cosh\zeta'\,\eta' - \sinh\zeta'\,\xi' \end{cases} \tag{3s}$$

and

$$\xi = -\sinh(\zeta' - \zeta)\,\eta' + \cosh(\zeta' - \zeta)\,\xi'$$
$$\eta = \cosh(\zeta' - \zeta)\,\eta' - \sinh(\zeta' - \zeta)\,\xi' \tag{4s}$$

The relative speed is $(d\xi' = 0)$

$$-\tanh(\zeta' - \zeta) = -\frac{\tanh\zeta' - \tanh\zeta}{1 - \tanh\zeta' \tanh\zeta} = \frac{\cos\varphi - \cos\varphi'}{1 - \cos\varphi\cos\varphi'} \tag{5s}$$

and the velocity transformation is

$$\frac{d\xi}{d\eta} = \frac{-\tanh(\zeta' - \zeta) + \frac{d\xi'}{d\eta'}}{1 - \tanh(\zeta' - \zeta)\frac{d\xi'}{d\eta'}} \tag{6s}$$

There is also a transformation between $\frac{d\xi}{d\eta} = \cos\varphi$ and $\frac{d\xi'}{d\eta'} = \cos\varphi'$. Let

$$\beta = -\tanh(\zeta' - \zeta), \gamma = 1/\sqrt{1 - \beta^2} \tag{7s}$$

We can easily derive the value of $\beta$ and $\gamma$ from reading the angle $\varphi$ and $\varphi'$ in Fig. 2s.

$$\beta = \frac{\cos\varphi - \cos\varphi'}{1 - \cos\varphi\cos\varphi'} \tag{8s}$$

$$\gamma = \frac{1 - \cos\varphi\cos\varphi'}{\sin\varphi\sin\varphi'} \tag{9s}$$

Therefore, we can treat $\eta$ as time label $t_M$ and $\xi$ as length label $x_M$ in the article. In this case, we have associated the $\tau(x, y)$ object with an inertial object with constant velocity. The two meta-chains are in straight lines (white dashed line) which make different angles to the SPP-rays. They correspond to two different inertial reference frames, observing the same inertial motion. When the trajectory intersects a contour line of $\tau$, it means the proper time observed by the object is $\tau$. Therefore, in the current picture, an event is represented by a line instead of a point in the conventional world-line diagram. A point in a conventional world-line diagram indicates $t_M$ and $x_M$ with the proper time obtained by integrating along a world-line. A point in our picture indicates a particular proper time $\tau$ and $l$, with $\Phi$ obtained by integrating the metric along a trajectory (Eq. (3)). We also note that the analogous (dimensionless) velocity here is defined as $v = dx_M/dt_M = (1/k_{\text{SPP}})d\Phi/dl = \cos\varphi$. It is the reciprocal of the velocity used in P. Genevet's work since we would like to associate to time-like instead of space-like events in relativity. $\zeta$ in Eq. (5) is the rapidity between the two observing frames. It is related to the velocities observed in the two frames by the velocity-subtraction formula in relativity. A non-zero rapidity signifies different observed velocities in the two frames, equivalently revealed as two non-parallel trajectories in the caustic picture. This transformation should include a simple case for the transformation between a straight trajectory perpendicular to the parallel SPP rays, which means $\varphi' = \pi/2$, and a straight trajectory that is not perpendicular. We can

easily write down:

$$d\eta' = d\tau \tag{10s}$$

In this case the transformation can be simplified as

$$\begin{cases} \eta = \dfrac{1}{\sin\varphi}(\eta' + \xi'\cos\varphi) \\ \xi = \dfrac{1}{\sin\varphi}(\xi' + \eta'\cos\varphi) \end{cases} \tag{11s}$$

Now, $\gamma = \dfrac{1}{\sin\varphi}$ and $\beta = \cos\varphi$. If $d\eta' = 0$, the length contraction $d\xi = d\xi'/\sin\varphi$. If $d\xi = 0$, the time dilation $d\eta = d\eta'\sin\varphi$. If we set $\Delta\varphi = \varphi - \varphi'$, we can draw the changes of $d\eta/d\eta'$ as in Fig. 3s(a) and $d\xi/d\xi'$ as in Fig. 3s(b), with increasing $\Delta\varphi$.

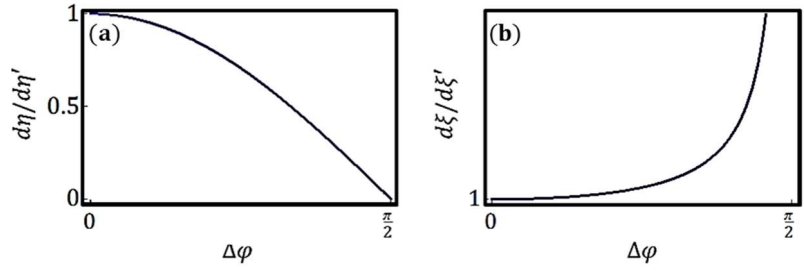

FIG. 3s, (a) the variation of $d\eta/d\eta'$ with change of $\Delta\varphi$. (b) the variation of $d\xi/d\xi'$ with change of $\Delta\varphi$.

## III. Transformation between constant movements and accelerations

In Fig. 2(a), we provide an example with two trajectories. The motion on one trajectory, marked by the white dashed line (in Fig. 4s it is black), has acceleration, and it has constant speed on the other trajectory marked by green dashed line. We draw the corresponding $\tau$ distribution in grayscale, with contours displayed by red lines. We calculate the $\tau(l)$ by defining $\tau(l) = \varphi(l) + \tan^{-1} dy/dx$. Thus, $l_1$, $l_2$, and the distribution, gradients and contour lines of $\tau$, are shown in Fig. 4s.

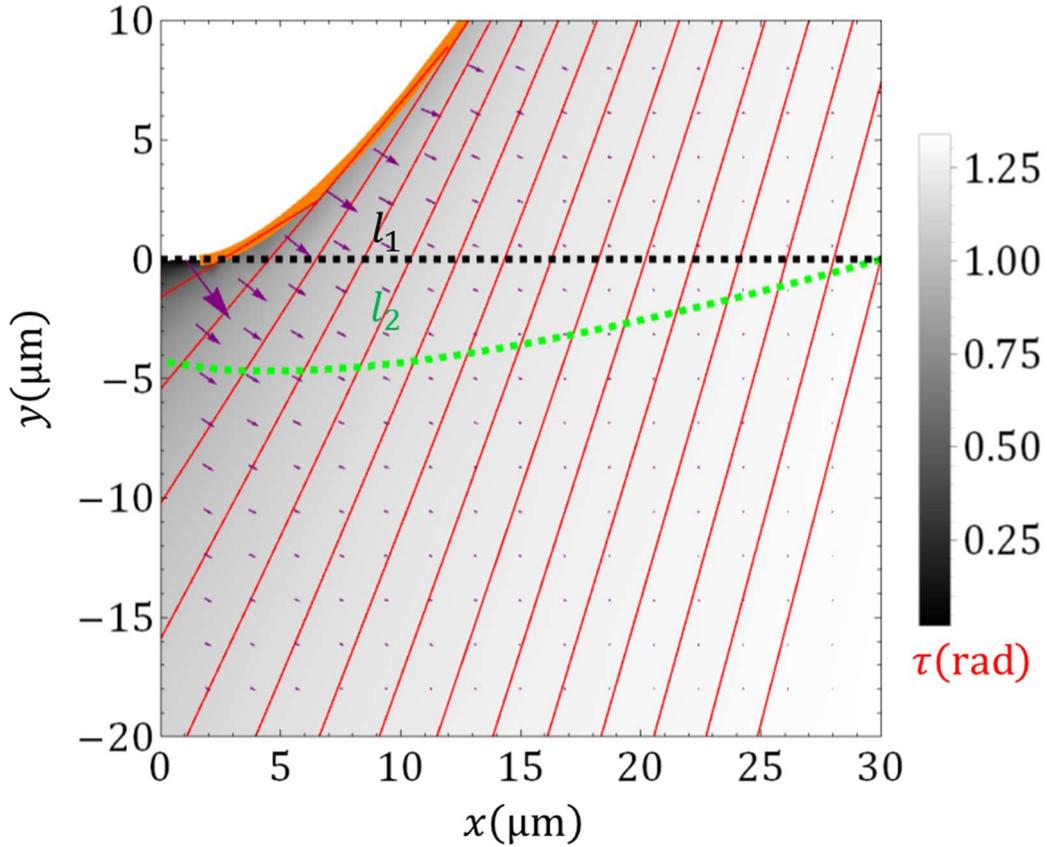

FIG. 4s. The distribution of $\tau$ is shown in grayscale, with purple arrows representing the distribution of $\nabla\tau$ and contour lines of the function $\tau$ marked in red. $l_1$ is the dashed line in black (in Fig. 2 it is white) and $l_2$ is the dashed line in green. These two lines intersect at point $(30\mu m, 0)$.

The white dashed horizontal line (in Fig. 4s it is black) has

$$\frac{dl_1}{d\Phi(l_1)/k_{spp}} = \frac{1}{v(l_1)} = -\frac{\sqrt{l_1}k_{spp}}{\sigma c_1} \tag{12s}$$

$$|\nabla\tau|(l_1) = -c_1\sigma/\left(2k_{spp}l_1^{\frac{3}{2}}(1 - \frac{c_1^2}{k_{spp}^2 l_1})\right) \tag{13s}$$

in which $c_1 = 2\sqrt{15}\pi/3$ and $\sigma = -1$ for incident spin. From this we know that when

$l_1 = \infty$, $\frac{d\Phi(l_1)/k_{spp}}{dl_1} = v(l_1) = 0$.

Since we have $\tau = \varphi(l) + \tan^{-1} dy/dx$, on the plane we can write

$$\frac{d\tau}{dl} = |\nabla\tau|(l) \sin(\tau - \tan^{-1}\frac{dy}{dx}) \tag{14s}$$

$$\frac{dl}{d\Phi/k_{spp}} = \frac{1}{\cos\left(\tau - \tan^{-1}\frac{dy}{dx}\right)} \tag{15s}$$

Apart from the metric-like discussion, if we confine ourselves to trajectories such that

$$q = \frac{d(\tau - p)}{d\tau} \tag{16s}$$

is a constant order parameter, where $p = \tan^{-1}\frac{dy}{dx}$, then a straight-line trajectory corresponds to $q = 1$ and the constant speed trajectory corresponds to $q = 0$. Then, we can write the geodesic equation for either trajectory as

$$\frac{d^2l}{d\tau^2} = \left(-\frac{d\ln|\nabla\tau|(l)}{dl}\left(\frac{dl}{d\tau}\right)^2 - c_{spp}q|\nabla\tau|(l)\frac{\frac{d\Phi}{k_{spp}}}{d\tau}\frac{dl}{d\tau}\right) \tag{17s}$$

$$= \left(-\frac{d\ln|\nabla\tau|(l)}{dl} - \frac{c_{spp}q|\nabla\tau|(l)}{v}\right)\left(\frac{dl}{d\tau}\right)^2$$

$$\frac{d^2\Phi/k_{spp}}{d\tau^2} = \left(\frac{d\Phi/k_{spp}}{dl}\frac{d^2l}{d\tau^2} + \frac{dl}{d\tau}\frac{d\frac{d\Phi/k_{spp}}{dl}}{d\tau}\right)$$

$$= \left(-\frac{d\ln|\nabla\tau|(l)}{d\Phi/k_{spp}} - c_{spp}q|\nabla\tau|(l) \right. \tag{18s}$$

$$\left. + \left(\frac{dl}{d\Phi/k_{spp}}\right)^2 \frac{d^2\Phi/k_{spp}}{dl^2}\right)\left(\frac{d\Phi/k_{spp}}{d\tau}\right)^2$$

for the first horizontal trajectory, with

$$\frac{d^2l_1}{d\tau^2} = \left(-\frac{d\ln g(l_1)}{dl_1} - \frac{c_{spp}g(l_1)}{v(l_1)}\right)\left(\frac{dl_1}{d\tau}\right)^2 = 2l_1\left(-2 + \frac{3\,k_{spp}^2 l_1}{c_1^2}\right) \tag{19s}$$

$$\frac{d^2\Phi(l_1)/k_{spp}}{d\tau^2}$$

$$= \left(-\frac{d\ln|\nabla\tau|(l_1)}{\frac{dl_1}{d\Phi/k_{spp}}dl_1} - \frac{c_{spp}|\nabla\tau|(l_1)}{\left(\frac{dl_1}{d\Phi/k_{spp}}\right)^2} + \frac{d^2\Phi/k_{spp}}{dl_1^2}\right)\left(\frac{dl_1}{d\tau}\right)^2 \tag{20s}$$

$$= \left(-\frac{2\,c_1\,l_1^{\frac{1}{2}}}{k_{spp}} + \frac{4\,k_{spp}l_1^{\frac{3}{2}}}{c_1}\right)$$

as its world line.

The green line has

$$v(l_2) \equiv 1/2 \tag{21s}$$

$$|\nabla\tau|(l_2) = \frac{d\left(\tan^{-1}\frac{dy_2}{dx_2}\right)/dl_2}{\sin(\cos^{-1}1/v)} \tag{22s}$$

Therefore,

$$|\nabla\tau|(l_2) = -\frac{1}{\sin\frac{2}{3}\pi}\frac{x_2''(l_2)}{\sqrt{1-x_2'(l_2)^2}} \tag{23s}$$

Similarly, for the second constant velocity trajectory, we have

$$\frac{d^2 l_2}{d\tau^2} = -\frac{d\ln|\nabla\tau|(l_2)}{dl_2}\left(\frac{dl_2}{d\tau}\right)^2 \tag{24s}$$

$$\begin{aligned}\frac{d^2\Phi(l_2)/k_{spp}}{d\tau^2} &= -\frac{d\ln|\nabla\tau|(l_2)}{\frac{dl_2}{d\Phi(l_2)/k_{spp}}dl_2}\left(\frac{dl_2}{d\tau}\right)^2 \\ &= -\frac{d\ln|\nabla\tau|(l_2)}{d\Phi(l_2)/k_{spp}}\left(\frac{d\Phi(l_2)/k_{spp}}{d\tau}\right)^2\end{aligned} \tag{25s}$$

as its world line.

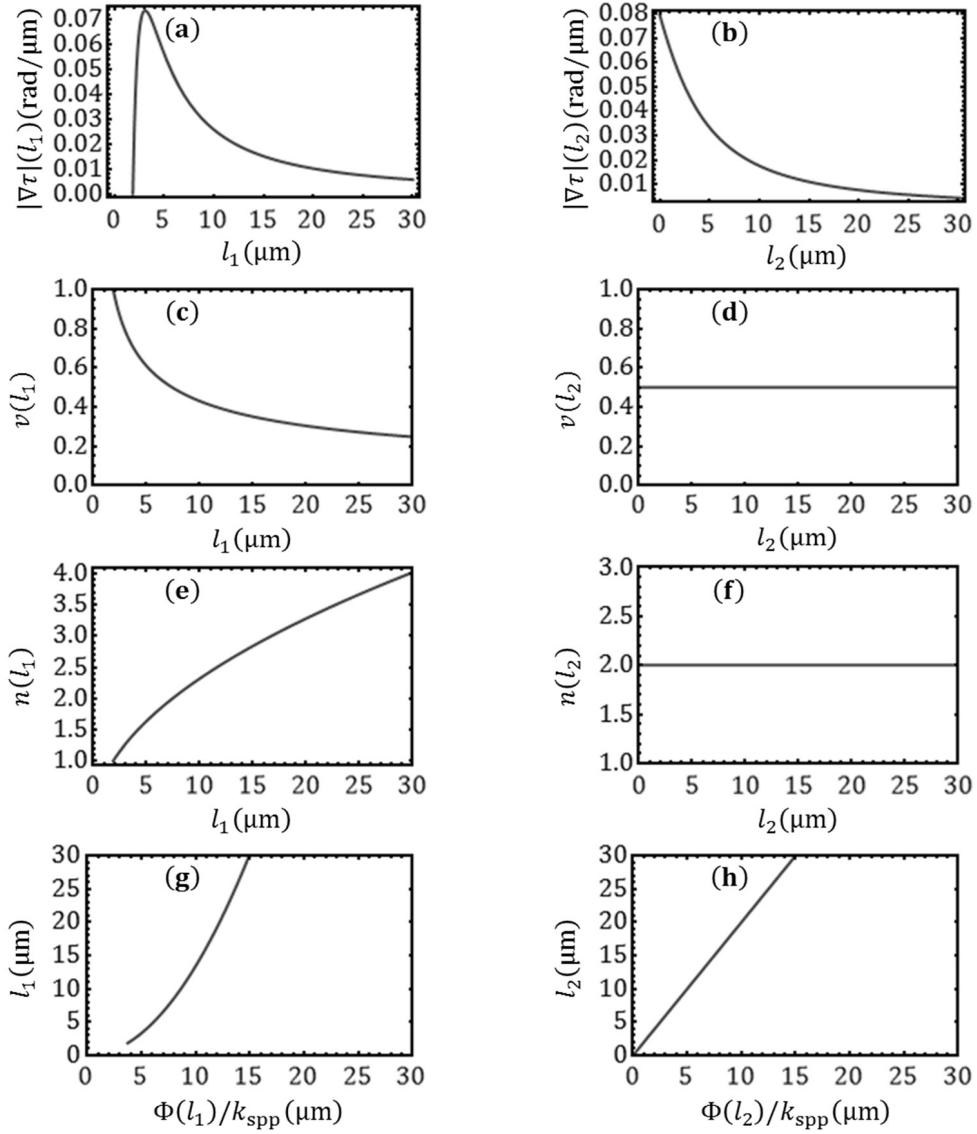

FIG. 5s. $l_1$ and $l_2$ should be read as time labels. (a, b) are numerical results with $\tau = \varphi + \tan^{-1} dy/dx$ of (a) $|\nabla\tau|(l_1)$, (b) $|\nabla\tau|(l_2)$. The velocity distribution of $l_1$ is (c) and of $l_2$ is (d). (e, f) are respectively calculated effective index $n$ on $l_1$ and $l_2$. (g, h) are respectively the world lines on $l_1$ and $l_2$.

Additionally, if we define the moving speed of photon along the meta-chain as $v = (d\Phi/dl)/k_{spp}$ based on gradient of geometry phase $\Phi$, we can define effective index of structure as $n = c/v = c \cdot k_{spp}/(d\Phi/dl)$.

In Fig. 5s, we numerically calculate Eqs. (12s, 13s, 21s, 22s) to derive Figs. 5s(a-d), and we acquire Figs. 5s(g, h) through Eqs. (17s, 18s, 24s, 25s). Based on definition of effective index of structure, we acquire Figs. 5s(e, f) for effective index $n$ on $l_1$ and $l_2$ respectively.

## IV. Transformation of Metrics

Consider the transformation $(x_C, t_C) \to (x'_C, t'_C)$. The metric should obey $g_{\mu\nu} = \frac{\partial x^i}{\partial x^\mu} \frac{\partial x^k}{\partial x^\nu} g_{ik}$. First we write down the metric,

$$g_{t_C t_C} = |\nabla \tau|^2 (l, \Phi)$$
$$g_{x_C x_C} = -|\nabla \tau|^2 (l, \Phi)$$
$$g_{t'_C t'_C} = |\nabla \tau|^2 (l', \Phi')$$
$$g_{x'_C x'_C} = -|\nabla \tau|^2 (l', \Phi')$$
(26s)

If we only consider the metric transformation with $\varphi(\tau) = \pm\varphi'(\tau) + n\pi, n \in$ Integer,

$$g_{t_C t_C} = |\nabla \tau|^2 (l, \Phi) = \frac{\partial t'_C}{\partial t_C} \frac{\partial t'_C}{\partial t_C} g_{t'_C t'_C} + \frac{\partial x'_C}{\partial t_C} \frac{\partial x'_C}{\partial t_C} g_{x'_C x'_C}$$

$$= |\nabla \tau|^2 (l', \Phi') \left( \frac{\partial t'_C}{\partial t_C} \frac{\partial t'_C}{\partial t_C} - \frac{\partial x'_C}{\partial t_C} \frac{\partial x'_C}{\partial t_C} \right)$$

$$= |\nabla \tau|^2 (l', \Phi') \frac{\partial t'_C}{\partial t_C} \frac{\partial t'_C}{\partial t_C} \left( 1 - \frac{\frac{\partial x'_C}{\partial t_C} \frac{\partial x'_C}{\partial t_C}}{\frac{\partial t'_C}{\partial t_C} \frac{\partial t'_C}{\partial t_C}} \right)$$
(27s)

$$= g_{t_C t_C} \frac{\sin^2 \varphi}{\sin^2 \varphi'} \frac{dt_C^2}{dt_C'^2} \frac{\partial t'_C}{\partial t_C} \frac{\partial t'_C}{\partial t_C} \left( 1 - \frac{\frac{\partial x'_C}{\partial t_C} \frac{\partial x'_C}{\partial t_C}}{\frac{\partial t'_C}{\partial t_C} \frac{\partial t'_C}{\partial t_C}} \right)$$

$$g_{x_C x_C} = -|\nabla \tau|^2 (l, \Phi) = \frac{\partial t'_C}{\partial x_C} \frac{\partial t'_C}{\partial x_C} g_{t'_C t'_C} + \frac{\partial x'_C}{\partial x_C} \frac{\partial x'_C}{\partial x_C} g_{x'_C x'_C}$$

$$= |\nabla \tau|^2 (l', \Phi') \left( \frac{\partial t'_C}{\partial x_C} \frac{\partial t'_C}{\partial x_C} - \frac{\partial x'_C}{\partial x_C} \frac{\partial x'_C}{\partial x_C} \right)$$

$$= |\nabla \tau|^2 (l', \Phi') \left( -\frac{\partial x'_C}{\partial x_C} \frac{\partial x'_C}{\partial x_C} \right) \left( 1 - \frac{\frac{\partial t'_C}{\partial x_C} \frac{\partial t'_C}{\partial x_C}}{\frac{\partial x'_C}{\partial x_C} \frac{\partial x'_C}{\partial x_C}} \right)$$
(28s)

$$= g_{x_C x_C} \frac{\tan^2 \varphi}{\tan^2 \varphi'} \frac{dx_C^2}{dx_C'^2} \left( \frac{\partial x'_C}{\partial x_C} \frac{\partial x'_C}{\partial x_C} \right) \left( 1 - \frac{\frac{\partial t'_C}{\partial x_C} \frac{\partial t'_C}{\partial x_C}}{\frac{\partial x'_C}{\partial x_C} \frac{\partial x'_C}{\partial x_C}} \right)$$

Thus, $\frac{\partial x'_C}{\partial t_C}$ and $\frac{\partial t'_C}{\partial x_C}$ should all be zero to maintain the equality, which means that $\frac{\partial x'_C}{\partial x_C} \frac{\partial x'_C}{\partial x_C} = \frac{\partial t'_C}{\partial t_C} \frac{\partial t'_C}{\partial t_C} = \frac{|\nabla \tau|^2 (l, \Phi)}{|\nabla \tau|^2 (l', \Phi')}$. Moreover, the transformation between frame $(x_C, t_C)$ and $(x'_C, t'_C)$ actually is a conformal transformation. The motion seen in $(x_C, t_C)$ has $\frac{dx_C}{dt_C} = \frac{dx'_C |\nabla \tau|(l', \Phi')/|\nabla \tau|(l, \Phi)}{dt'_C |\nabla \tau|(l', \Phi')/|\nabla \tau|(l, \Phi)} = \frac{dx'_C}{dt'_C} = \cos \varphi' = \cos \varphi$, where after transformation the velocity equals the definition.

# V. The Rindler-Analogue Transformation

In analogy to the Rindler transformation in the 2D situation, we have

$$d\tau^2 = \beta^2 e^{2\alpha t}(dt^2 - dx^2) \tag{29s}$$

in which $\alpha$ and $\beta$ are constants. In our platform, a two-dimensional plane, we have

$$d\tau^2 = |\nabla\tau|(l)^2 \left(dl^2 - d\Phi^2/k_{spp}^2\right) \tag{30s}$$

When we set $|\nabla\tau|(l) = \beta e^{\alpha l}$, and assume this trajectory is horizontal ($l = x$),

$$\frac{d\tau}{dx} = |\nabla\tau|(x)\sin\varphi(x) \tag{31s}$$

Setting the value of $\tau = \varphi + \tan^{-1} dy/dx$,

$$\tau = f(\varphi) \tag{32s}$$

We write

$$\frac{df}{d\varphi}\frac{d\varphi}{dx} = |\nabla\tau|(x)\sin\varphi(x) \tag{33s}$$

Solving Eq. (33s) with $|\nabla\tau|(l) = \beta e^{\alpha l}$ and $\tau = \varphi$, we have

$$\tau(x) = 2\cot^{-1} e^{-\frac{\beta e^{\alpha x}}{\alpha}} + Constant \tag{34s}$$

$$\frac{dx}{d\Phi/k_{spp}} = \frac{1}{\cos\varphi(x)} = -\coth\left(\frac{\beta e^{\alpha x}}{\alpha}\right) \tag{35s}$$

in which $\beta$ and $\alpha$ are constants and $\beta$ is a scale constant. Notice that $dx/(d\Phi/k_{spp})$ determines the slope of the SPP ray emitting from the trajectory, so that $|\nabla\tau|(l)$ will give the value difference on the SPP ray ($\tau$ line) between different slopes.

Therefore, for simplification, we do not define $|\nabla\tau|(l_1)/|\nabla\tau|(l_2) = \beta e^{\alpha x}$, but instead set $\nabla(l_2) = Constant$ and solve $\frac{d\tau}{dx} = |\nabla\tau|(x)\sin\varphi(x)$ using Eqs. (34s,35s) above by choosing the gauge $\tau = \angle\hat{t}$, which means $\varphi(x) = \tau(x)$ on $l_1$. In the experiment, we chose $\alpha = 1/6, \beta = 1/60$ for our design and one micron for the length unit. For the green line $l_3$ in Fig. 4(a), $v_3 = (d\Phi/k_{spp})/dl_3 \equiv -1/2$ and $\varphi_3 \equiv 120°$.

We show the corresponding $\tau$ in grayscale, with $\nabla\tau$ in purple and the contour lines of the function $\tau$ in red in Fig. 6s, choosing $\tau = \varphi(l) + \tan^{-1} dy/dx$.

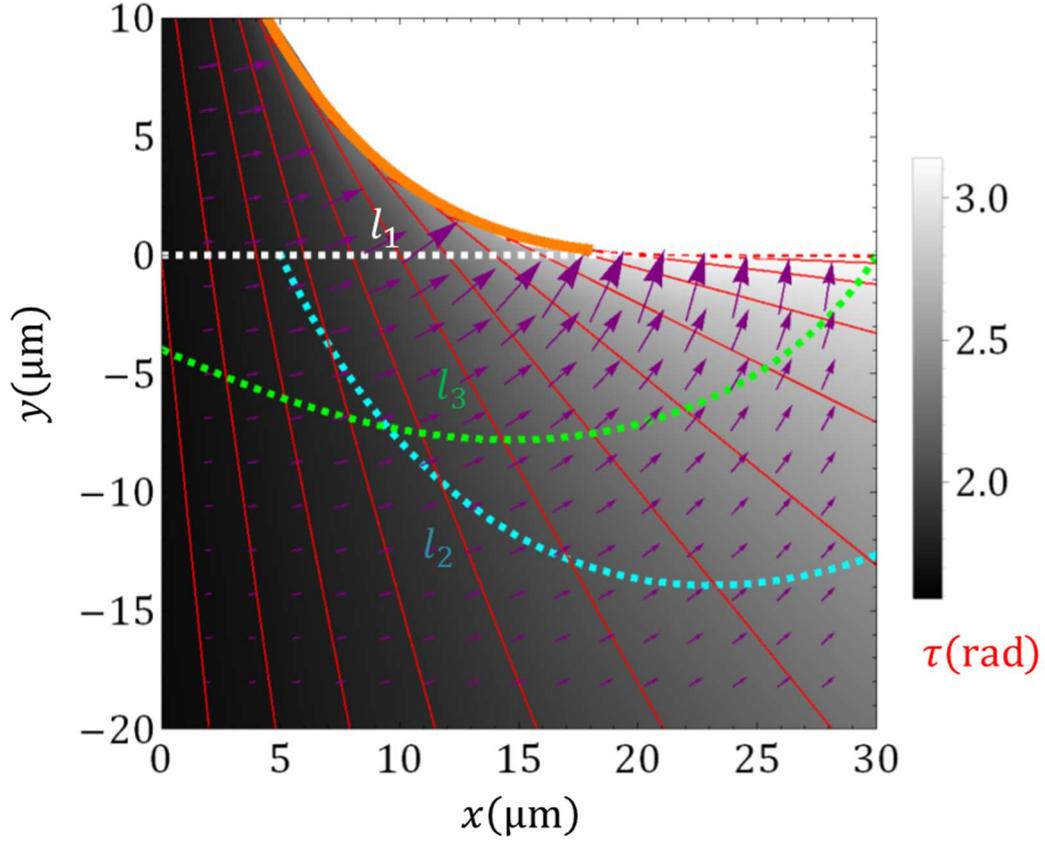

FIG. 6s. The distribution of $\tau$ is shown in grayscale with purple arrows representing the distribution of $\nabla\tau$ and the contour lines of the function $\tau$ marked in red. Due to rapid changes of $\nabla\tau$ in the top right corner, the accuracy of the amplitude there is little bit low because of the interpolation. Also, we mark $l_1$ in white, $l_2$ in cyan and $l_3$ in green.

Since we determine the $\tau$ plane in this case, we can numerically derive the corresponding $|\nabla\tau|(l)$ shown in Figs. 7s(a-c). Also, $v(l)$ for the three trajectories is shown in Figs. 7s(d-f), and, similar to the second part of the supplement, we draw effective index in Figs. 7s(g-i) and world lines with $l$ being time in Figs. 7s(j-l).

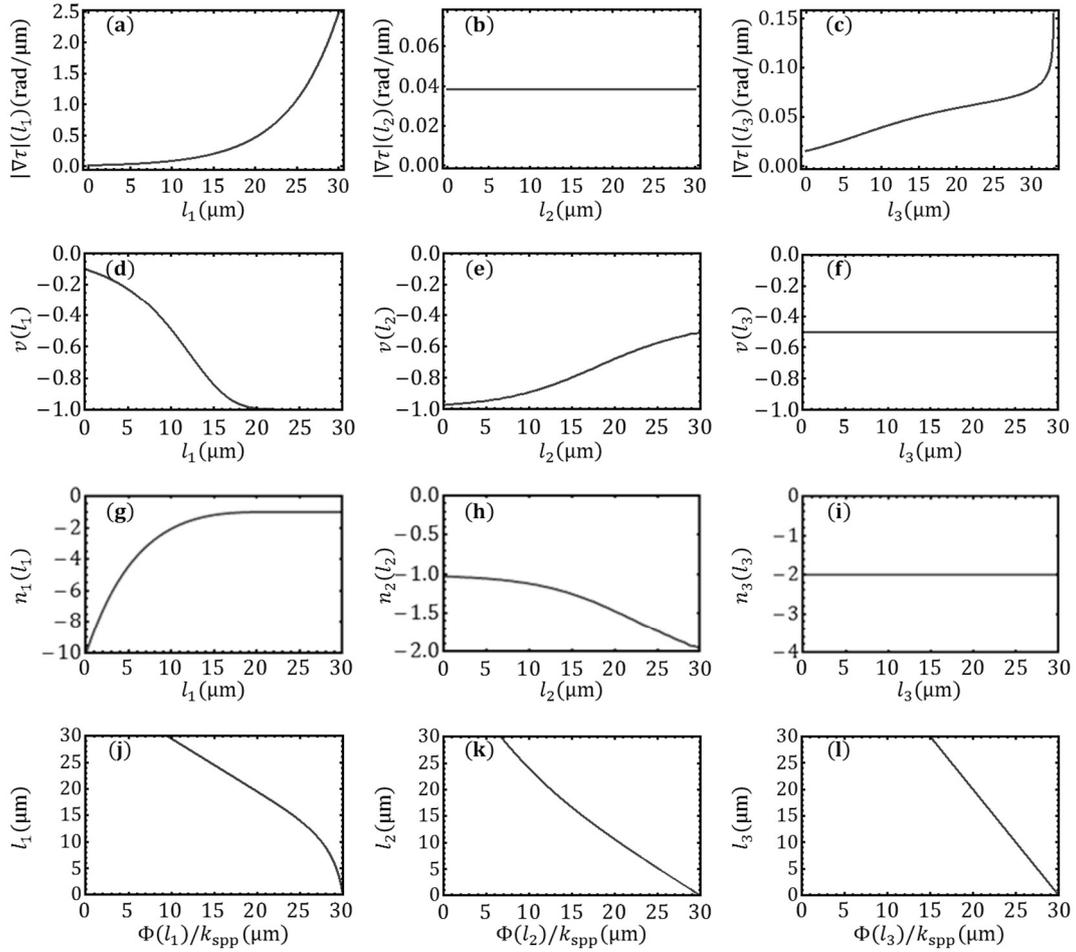

FIG. 7s, the calculated (a) $|\nabla \tau|(l_1)$, (d) $v(l_1)$, (j) world line for motion on $l_1$; the calculated (b) $|\nabla \tau|(l_2)$, (e) $v(l_2)$, (k) world line for motion on $l_2$; the calculated (c) $|\nabla \tau|(l_3)$, (f) $v(l_3)$, (l) world line for motion on $l_3$. The three world lines show the motion is directed backwards. The effective index $n$ on $l_1$ is shown in (g); $n$ on $l_2$ is shown in (h); $n$ on $l_3$ is shown in (i).

## VI. Gauge transformations

In implementing Bremsstrahlung radiation on the metasurface, we purposely set a value of $\tau$ equaling the angle between the SPP ray and $x$-direction. However, as mentioned in the article, there is a flexibility when choosing the gauge while still retaining the ratio between different $g(l)$ which actually describes the relative properties between the coordinates. Here, we show another choice $\tilde{\tau}$ for setting the value on the $\tau$ line equaling the $x$-position of the intersection point of the $\tau$ line and $x$ axis.

In the first case, the relation between the two gauges is $\tilde{\tau} = \left(\frac{c_1}{k_{spp}\cos\tau}\right)^2$ where $c_1 = 2\sqrt{15}\pi/3$ is a constant defined in the second part of the supplement. The comparison of properties is shown in Fig. 8s. We can see that when $\tau$ changes, $\nabla\tau$ will also change (in Fig. 8s(b) compared to Fig. 8s(a)), and $|\nabla\tau|(l_i) \neq |\nabla\tilde{\tau}|(l_i)$ for $i = 1,2$ in Figs. 8s(c-f), while for most important properties we still have $\frac{g(l_1)}{g(l_2)} = \frac{\tilde{g}(l_1)}{\tilde{g}(l_2)}$. We show $\frac{|\nabla\tilde{\tau}|(l_1)}{|\nabla\tilde{\tau}|(l_2)} = \frac{f'(\tau)|\nabla\tau|(l_1)}{f'(\tau)|\nabla\tau|(l_2)} = \frac{|\nabla\tau|(l_1)}{|\nabla\tau|(l_2)}$ in Figs. 8s(g, h), in accordance with the length transformation for $\tau(l_1) = \tau(l_2)$ and $\tilde{\tau}(l_1) = \tilde{\tau}(l_2)$ which determines the relationship between the two spaces.

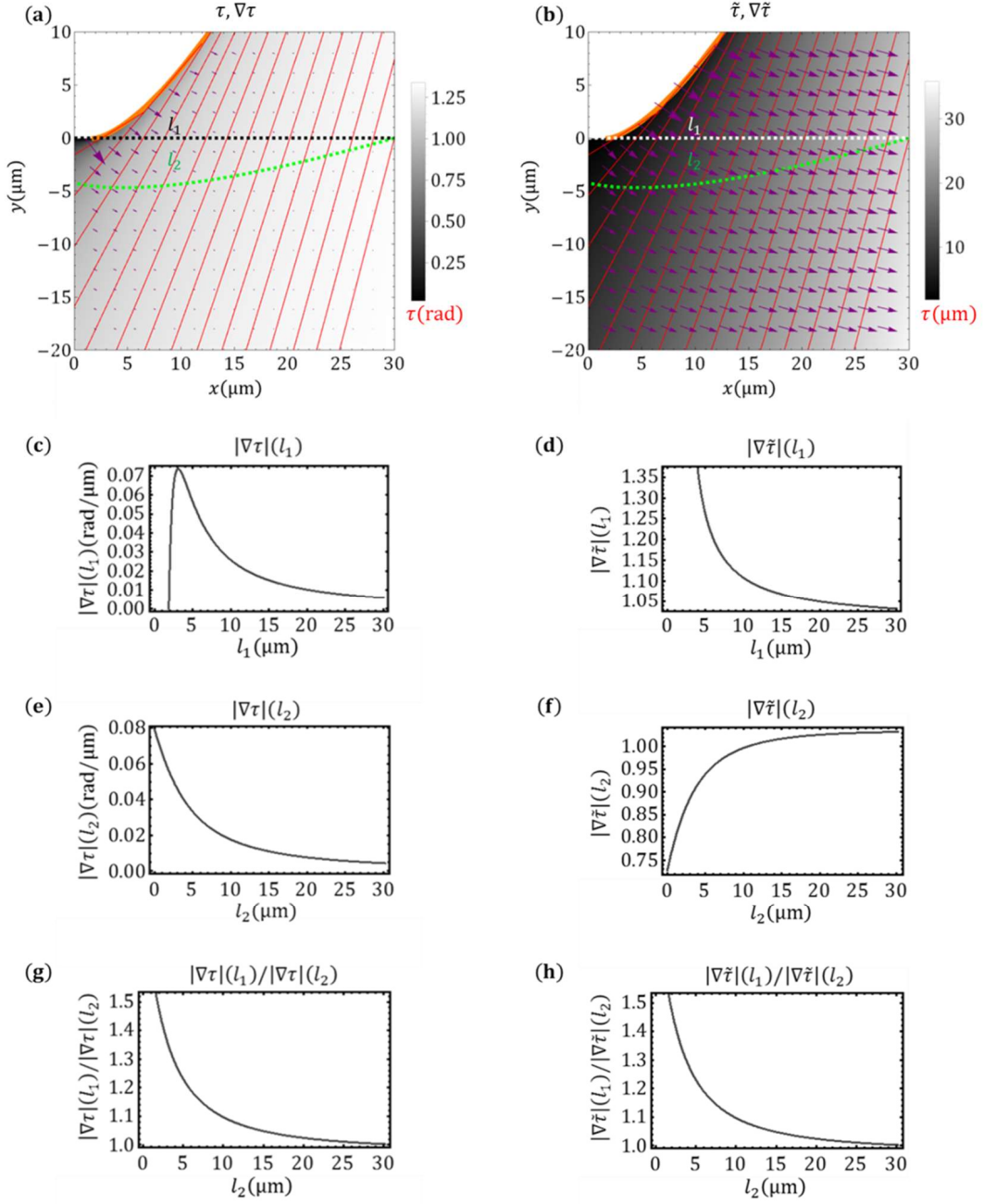

FIG. 8s. $l_1 = 0$ corresponding to $x = y = 0$; $l_2 = 0$ corresponding to $x = 0.32\mu m, y = -4.29\mu m$. $l_1$ and $l_2$ share the same point $(30\mu m, 0)$. (a) $\tau$ plane with $\nabla\tau$ in purple, contour lines of $\tau$ in red, black dashed line $l_1$ and green dashed line $l_2$. (b) $\tilde{\tau}$ plane with $\nabla\tilde{\tau}$ in purple, contour lines of $\tilde{\tau}$ in red, white dashed line $l_1$ and green dashed line $l_2$. The ratio (g) between (c)$|\nabla\tau|(l_1)$ and (e)$|\nabla\tau|(l_2)$ according to $\tau(l_1) = \tau(l_2)$, is the same as the ratio (h) between (d)$|\nabla\tilde{\tau}|(l_1)$ and (f)$|\nabla\tilde{\tau}|(l_2)$ according to $\tilde{\tau}(l_1) = \tilde{\tau}(l_2)$.

In the Rindler-analogous case, the relation between two gauges is $\tilde{\tau} = \frac{1}{\alpha}\ln\left(\frac{\alpha}{\beta}\coth^{-1}\frac{-1}{\cos\tau}\right)$. We choose $l_1$ and $l_2$ out of the three trajectories to show the comparison of properties shown in Fig. 9s. Comparing the left column to the right, we can see that when $\tau$ changes, $\nabla\tau$ will also change in Fig. 9s(b) compared to Fig. 9s(a).

Additionally, $g(l_i) \neq \tilde{g}(l_i)$ for $i = 1,2$ in Figs. 9s(c-f), while for the most important property we still have $\frac{g(l_1)}{g(l_2)} = \frac{\tilde{g}(l_1)}{\tilde{g}(l_2)}$ in Figs. 9s(g, h), in accordance with the length transformation for $\tau(l_1) = \tau(l_2)$ and $\tilde{\tau}(l_1) = \tilde{\tau}(l_2)$ which determines the relationship between the two spaces.

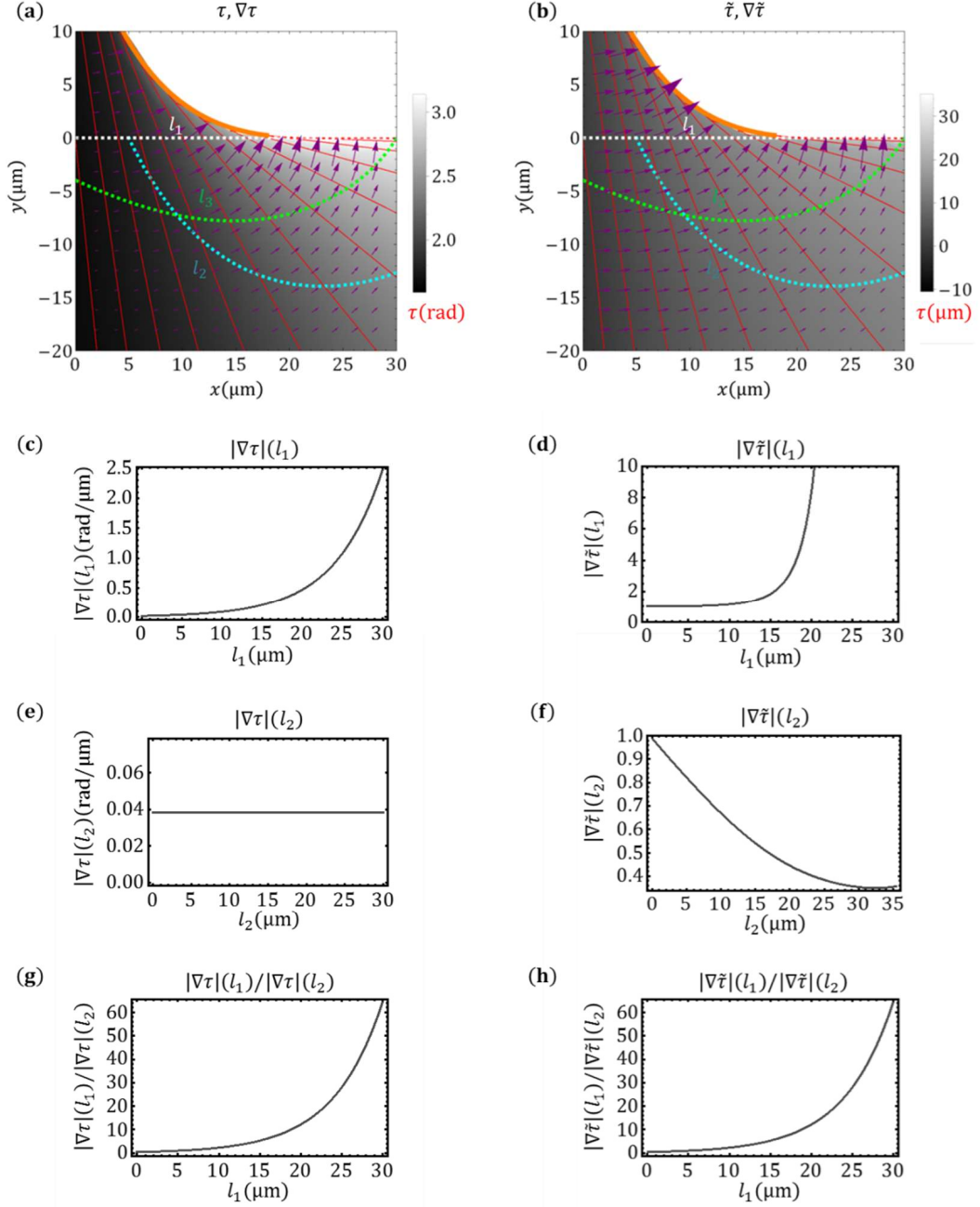

FIG. 9s. $l_1 = 0$ corresponds to $x = y = 0$; $l_2 = 0$ corresponds to $x = 5.50\mu m$, $y = -1.00\mu m$.; $l_3 = 0$ corresponds to $x = 0.14\mu m$ and $y = -4.02\mu m$. $l_1$ and $l_3$ share the same point $(30\mu m, 0)$. (a) $\tau$ plane with $\nabla\tau$ in purple, the contour lines of $\tau$ in red, white dashed line $l_1$, cyan dashed line $l_2$ and green dashed line $l_3$. (b) $\tilde{\tau}$ plane with $\nabla\tilde{\tau}$ in purple, the contour lines of $\tilde{\tau}$ in red, white dashed line $l_1$, cyan dashed line $l_2$ and green dashed line $l_3$. The ratio (g)

between (c) $|\nabla\tau|(l_1)$ and (e) $|\nabla\tau|(l_2)$ obeying $\tau(l_1) = \tau(l_2)$ is the same as the ratio (h) between (d) $|\nabla\tilde{\tau}|(l_1)$ and (f) $|\nabla\tilde{\tau}|(l_2)$ obeying $\tilde{\tau}(l_1) = \tilde{\tau}(l_2)$.

## VII. Contribution of geometric phase in our scheme

In our model, the nano-slot can be regarded as an electric dipole $p = \frac{1}{2}\left((t_u + t_v)(\hat{x} \pm i\hat{y}) + (t_u - t_v)e^{\pm 2i}\ (\hat{x} \mp i\hat{y})\right)$ (as in Nat. Commun. 6, 8360 (2015)). In the dipole excitation formula, we have two terms: co-polarized term $p_1 = \frac{1}{2}\left((t_u + t_v)(\hat{x} \pm i\hat{y})\right)$ and cross-polarized term $p_2 = \frac{1}{2}\left((t_u - t_v)e^{\pm 2i\theta}(\hat{x} \mp i\hat{y})\right)$. Here, only the cross-polarized term carries spin-dependent geometry phase $e^{\pm 2i\theta}$, and the co-polarized term has no geometric phase. Here, we simulate a simple case that nano-slot angle $\theta = -3\pi/5 + \pi x/(3\mu m)$ as shown in Fig. 10s.

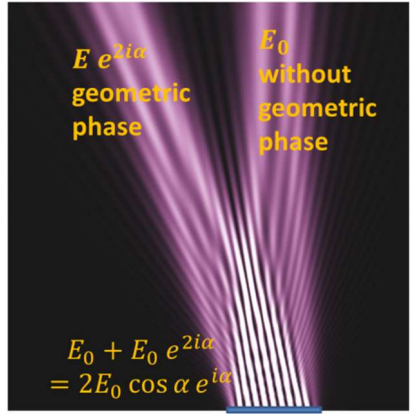

FIG. 10s. Simulate generation of surface plasmons. Excite about 50 nano-slots located on blue rectangle area which is on the bottom of simulation area and $t_u = 1, t_v = 0$. The result provides radiation with geometric phase, radiation without geometric phase and their interference part.